\documentclass[conference,a4paper]{APSIPA2025}
\usepackage{amsmath}
\usepackage{graphicx}
\usepackage{multirow}
\usepackage{threeparttable}
\usepackage[backend=biber,style=ieee,]{biblatex}
\usepackage{geometry}
\usepackage{fancyhdr}

\fancypagestyle{firststyle}{
  \fancyhf{}
  \fancyhead[C]{2025 Asia Pacific Signal and Information Processing Association Annual Summit and Conference (APSIPA ASC)}
}
\usepackage{booktabs}
\usepackage{multirow}
\usepackage{amsfonts}
\usepackage[linesnumbered,ruled,vlined]{algorithm2e}
\SetAlgorithmName{Algorithm}{Algorithm}{Algorithm}
\newcommand{\AlgoFontSize}{\small}

\begin{document}

\title{Peransformer: Improving Low-informed Expressive Performance Rendering with Score-aware Discriminator}

\author{
\authorblockN{
Xian He\authorrefmark{1}, Wei Zeng\authorrefmark{1}\authorrefmark{2}, and
Ye Wang\authorrefmark{1}\authorrefmark{2}\authorrefmark{3}
}

\authorblockA{
\authorrefmark{1}
School of Computing, National University of Singapore \\
\authorrefmark{2}
Integrative Sciences and Engineering Programme, NUS Graduate School \\
E-mail: \{xian.he, w.zeng\}@u.nus.edu.sg\\
\authorrefmark{3}Corresponding author. Email: wangye@comp.nus.edu.sg}
}

\maketitle
\thispagestyle{firststyle}
\pagestyle{fancy}

\begin{abstract}

Highly-informed Expressive Performance Rendering (EPR) systems transform music scores with rich musical annotations into human-like expressive performance MIDI files. While these systems have achieved promising results, the availability of detailed music scores is limited compared to MIDI files and are less flexible to work with using a digital audio workstation (DAW). Recent advancements in low-informed EPR systems offer a more accessible alternative by directly utilizing score-derived MIDI as input, but these systems often exhibit suboptimal performance. Meanwhile, existing works are evaluated with diverse automatic metrics and data formats, hindering direct objective comparisons between EPR systems. In this study, we introduce Peransformer, a transformer-based low-informed EPR system designed to bridge the gap between low-informed and highly-informed EPR systems. Our approach incorporates a score-aware discriminator that leverages the underlying score-derived MIDI files and is trained on a score-to-performance paired, note-to-note aligned MIDI dataset. Experimental results demonstrate that Peransformer achieves state-of-the-art performance among low-informed systems, as validated by subjective evaluations. Furthermore, we extend existing automatic evaluation metrics for EPR systems and introduce generalized EPR metrics (GEM), enabling more direct, accurate, and reliable comparisons across EPR systems. The repository is available at https://github.com/Bigstool/peransformer.

\end{abstract}

\section{Introduction}\label{sec:introduction}

Musical expressivity arises not only from composition but also from the performer’s interpretation and delivery. Nuances in tempo, dynamics, articulation, and other performance parameters critically influence how a piece resonates with listeners \cite{carlos2018computational}. Expressive Performance Rendering (EPR) refers to the computational task of rendering performances that mimic human expressiveness from existing music compositions. EPR systems can serve as virtual reference performers in music education, function as plugins to streamline music production, and operate as surrogates for human performance in tasks such as Audio-to-Score Transcription (A2S) \cite{zeng2024end}, particularly in scenarios where human-recorded data is limited.

EPR systems are generally categorized by their input modalities: highly-informed systems leverage score-level information such as tempo, time signature, key signature, and note value, whereas low-informed systems operate on more accessible note-level information like onset, offset, and velocity. While highly-informed systems have achieved convincing results \cite{chacon2016basis, jeong2019virtuosonet, jeong2019graph, borovik2023scoreperformer}, their flexibility and accessibility are inherently constrained — score notations (e.g., MusicXML) are not only harder to obtain but also more cumbersome to manipulate compared to MIDI-based representations, especially when working with a digital audio workstation (DAW). Recent advances in low-informed EPR systems \cite{renault2023expressive, tang2023reconstructing} address this issue by directly using score-derived MIDI as input, offering greater practicality. However, subjective evaluations have shown that their expressive quality still falls short compared to highly-informed models. Additionally, many of these systems are trained on unpaired \cite{renault2023expressive} or pseudo-paired \cite{tang2023reconstructing} data, limiting their ability to render performances coherent with the underlying composition.

Another challenge is the evaluation of EPR models using automatic metrics. Various metrics, such as mean squared error (MSE) \cite{jeong2019virtuosonet, jeong2019graph, rhyu2022sketching}, Pearson correlation coefficient \cite{jeong2019virtuosonet, jeong2019graph, rhyu2022sketching}, and mean absolute error (MAE) \cite{tang2023reconstructing}, have been employed for this purpose. However, current evaluation procedures apply these metrics directly to raw model outputs. Converting model outputs into the standard MIDI format often disrupts input-target alignment due to expressive variations in note ordering, making automatic evaluation non-trivial. Furthermore, differences in data representations and evaluation workflows across studies make direct comparisons between results difficult.

To address these limitations, we propose Peransformer, a low-informed expressive performance rendering (EPR) model based on the Transformer encoder \cite{vaswani2017attention}. In order to encourage the renditions to remain faithful to the composition, we utilize a score-aware discriminator, which conditions on MIDI files derived from the original score. We construct ASAP-MIDI, a note-to-note aligned, score-to-performance paired dataset with the ASAP dataset \cite{foscarin2020asap} and a alignment tool \cite{nakamura2017performance} for the training of Peransformer. Subjective listening tests demonstrate that our model significantly outperforms existing low-informed EPR approaches and greatly narrows the performance quality gap between low-informed and high-informed models.

Furthermore, we introduce Generalized EPR Metrics (GEM) - a set of evaluation tools that standardize both data formatting and metric computation. GEM facilitates consistent evaluation by using alignment tools \cite{nakamura2017performance}, enabling comparisons across any EPR model that outputs MIDI data.

In summary, our contributions are threefold:

\begin{enumerate}
    \item Peransformer – a novel low-informed EPR model with a score-aware discriminator conditioned on the original composition.
    \item ASAP-MIDI - an open access, score-to-performance paired, note-to-note aligned dataset for training low-informed EPR models.
    \item Generalized EPR Metrics (GEM) – a standardized evaluation workflow that enables direct,  accurate, and reliable comparisons between EPR models.
\end{enumerate}

\section{Methodology}

\subsection{Dataset}

We source the score and performance MIDI files of ASAP-MIDI from the Aligned Scores and Performances (ASAP) dataset \cite{foscarin2020asap}, a classical piano music dataset. However, ASAP only provides beat-level alignment. We obtain score-to-performance, note-to-note alignment using the alignment tool proposed by Nakamura et al. \cite{nakamura2017performance}. Each composition has at least one human performance.

As a quality control measure, a performance MIDI is discarded if more than 6\% of the notes in both the score and the performance MIDI failed to match, or if more than 3\% of the pairs of matched notes have different pitches. The thresholds are empirically chosen so that no obvious difference between performances before and after alignment is perceived, and approximately 90\% of the human performances in the dataset are retained.

For every performance, we further scale the score MIDI to match its length. Finally, the dataset is split with a rough ratio of 8:1:1. The statistics of the processed dataset are shown in Table \ref{tab:dataset}.

\begin{table}[]
\centering
\begin{tabular}{@{}ccccc@{}}
\toprule
Split & Composers & \begin{tabular}[c]{@{}c@{}}Compo-\\ sitions\end{tabular} & \begin{tabular}[c]{@{}c@{}}Score\\ Notes\end{tabular} & \begin{tabular}[c]{@{}c@{}}Performance\\ Duration\end{tabular} \\ \midrule
Train & 14        & 168          & 441k        & 56.6h                \\
Val   & 9         & 22           & 62.5k       & 11.2h                \\
Test  & 7         & 20           & 53.0k       & 6.16h                \\ \midrule
Total & 15        & 210          & 556k        & 74.0h                \\ \bottomrule
\end{tabular}
\caption{Statistics of the aligned and split dataset.}
\label{tab:dataset}
\end{table}

\subsection{Data Representation}\label{sec:datarepresentation}

We adopted the encoding method for model input in \cite{renault2023expressive} and applied it to both the score and performance MIDI. A performance $\boldsymbol{p} = (n_1, n_2, n_3, ...)$ is defined as a list of notes, where the j-th note $n_j = (i_j, d_j, p_j, v_j)$ is a tuple of the Inter-Onset Interval (IOI) in seconds $i \in \mathbb{R}_{\geq 0}$, duration in seconds $d \in \mathbb{R}_{\geq 0}$, MIDI note number $p \in \{m \in \mathbb{Z} \mid 0 \leq m \leq 127\}$, and MIDI velocity $v \in \{m \in \mathbb{Z} \mid 0 \leq m \leq 127\}$. Specifically, $i_j$ is calculated as:

\begin{equation}
    i_j =
\begin{cases}
0 & \text{if } j = 1 \\
o_j - o_{j-1} & \text{if } j = 2, 3, ..., m
\end{cases}\text{,}
\end{equation}
and $d_i$ is calculated as:

\begin{equation}
    d_j = f_j - o_j \text{  for } j = 1, 2, ..., m\text{,}
\end{equation}
where $o$ is the absolute onset time and $f$ is the absolute offset time, in seconds. We further apply z-score scaling to the four features as a standardization measure.

\begin{figure*}
\centering
\includegraphics[width=0.6\linewidth]{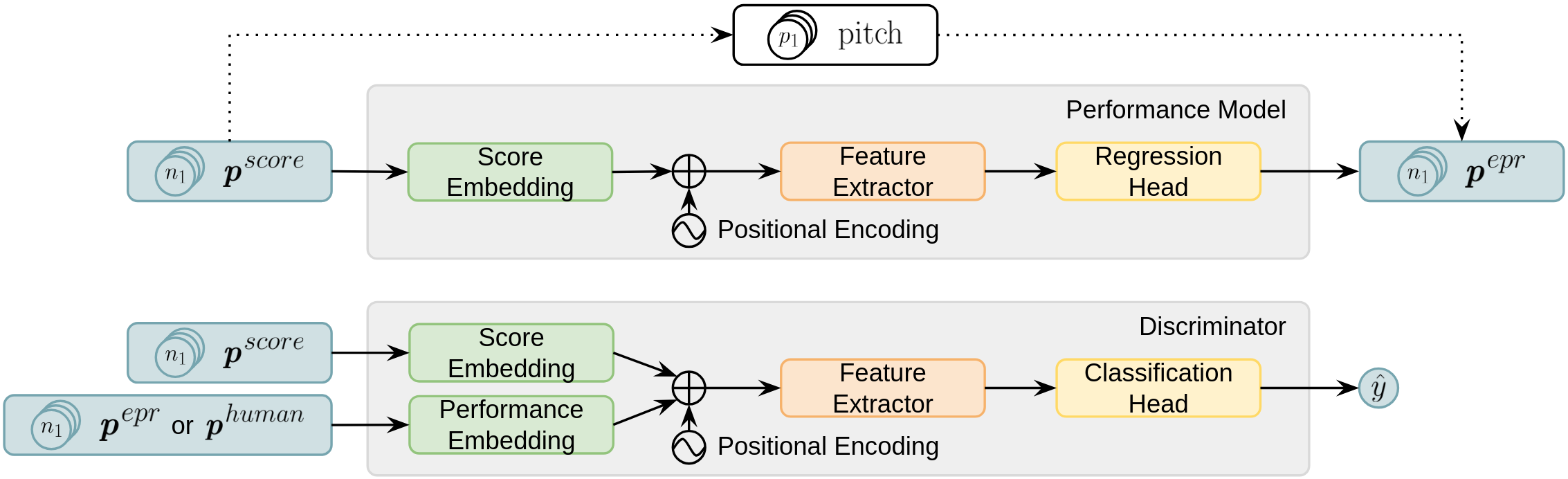}
\caption{The performance model and the discriminator.}
\label{fig:model}
\end{figure*}

\subsection{Model Architecture}

\subsubsection{Performance Model}

The performance model is comprised of a score embedding layer, a feature extractor layer, and a regression head, as shown in Figure \ref{fig:model}. The score embedding layer is a fully connected layer that transforms the dimensions of the input $\boldsymbol{p}^{s}$ to the dimensions of the feature extractor. A sinusoidal positional encoding from \cite{vaswani2017attention} is added to the score embedding before being passed into the feature extractor, which is a stack of transformer encoder blocks \cite{vaswani2017attention}. The regression head is another fully connected layer with an output dimension of 3, corresponding to the predicted IOI, duration, and velocity. Since the pitchs of the notes are unchanged during this process, they are copied from the input $\boldsymbol{p}^{s}$ and assembled with the predicted values as the rendition $\boldsymbol{p}^{e}$.

\subsubsection{Discriminator}

The score-aware discriminator takes either the rendition $\boldsymbol{p}^{\text{e}}$ or the human performance $\boldsymbol{p}^{\text{h}}$ as input through the performance embedding layer, while conditioning on the score MIDI $\boldsymbol{p}^{\text{s}}$ passed through the score embedding. Also shown in Figure \ref{fig:model}, all layers share the same architecture as the performance model, with the difference that the two embeddings are added together with the positional embedding, and the output dimension of the classification head is 1, producing the classification logit.

\subsubsection{Loss Functions}

The loss function of the performance model $\Psi ^{\text{P}}$ given the discriminator $\Psi ^{\text{D}}$ is formulated as:

\begin{multline}
    L_{\Psi ^{\text{P}}} = \dfrac{\lambda _0 \text{BCE}(\Psi ^{\text{D}}(\boldsymbol{p}^{\text{s}}, \boldsymbol{p}^{\text{e}}))}{c} + \dfrac{\lambda _1 \text{MSE}(\boldsymbol{i}^{\text{h}}, \boldsymbol{i}^{\text{e}})}{c} + \\ \dfrac{\lambda _2 \text{MSE}(\boldsymbol{d}^{\text{h}}, \boldsymbol{d}^{\text{e}})}{c} + \dfrac{\lambda _3 \text{MSE}(\boldsymbol{v}^{\text{h}}, \boldsymbol{v}^{\text{e}})}{c}\text{,}
\end{multline}
where $\text{BCE}()$ is the binary cross-entropy loss. $\text{MSE}()$ is the mean squared error loss, which is added to stabilize training \cite{isola2017image}. $\boldsymbol{i}$, $\boldsymbol{d}$, $\boldsymbol{v}$, are the lists of IOI, duration, and velocity from the corresponding performance. $\lambda$ are the weight balancing terms. $c$ is the count of human performances of the composition in the dataset, which balances the highly skewed dataset with various numbers of performances for each composition.

The loss of the discriminator is formulated as:

\begin{equation}
    L_{\Psi ^{\text{D}}} = \dfrac{\lambda _4 \text{BCE}(\Psi ^{\text{D}}(\boldsymbol{p}^{\text{s}}, \boldsymbol{p}^{\text{h}}))}{c} + \dfrac{\lambda _5 \text{BCE}(1 - \Psi ^{\text{D}}(\boldsymbol{p}^{\text{s}}, \boldsymbol{p}^{\text{e}}))}{c}\text{.}
\end{equation}

\subsection{The Generalized EPR Metrics (GEM)}

Following commonly used automatic metrics for the evaluation of EPR models, we calculate the mean squared error (MSE) and Pearson correlation coefficient between the rendition and the human performance over the IOI, duration, and velocity features. To allow for MIDI-based evaluation, hence enabling the application of GEM on any EPR model with MIDI output, we apply the aforementioned alignment tool \cite{nakamura2017performance} on the rendition and the human performance to recover the note-to-note correspondence which may be perturbed during the conversion from the raw model output to MIDI. Since each composition in the dataset can have multiple human performances, drawing inspiration from ROUGE \cite{lin2004rouge}, an established metric for text summarization, we treat multiple performances of the same composition as multiple references. One result is calculated between the rendition and each reference, and the best result is kept to represent the performance of the model on the given composition.

To enhance the reproducibility of GEM, we proceed to define a standardized evaluation process and a unified data format, starting from the following preprocessing steps:

\begin{itemize}
    \item Align the renditions to each of the human performances of the corresponding composition using \cite{nakamura2017performance}.
    \item Convert the aligned MIDI files to the data representation defined in \ref{sec:datarepresentation}, but without z-score scaling to preserve all data in the original units.
\end{itemize}

Then, without loss of generality, the application of GEM on the duration feature is given in Algorithm \ref{algo:metric}. The same can also be applied to the IOI and velocity predictions.

\begin{algorithm}
\AlgoFontSize{\small}
\caption{Duration Metric Calculation}
\label{algo:metric}
\KwIn{\texttt{comps\_e}, \texttt{comps\_h}: Lists of compositions by EPR model or human, each composition is a list of performances}
\KwOut{\texttt{l}: MSE, \texttt{p}: Pearson correlation}

Initialize \texttt{L\_sum, L\_n, P\_sum} $\gets 0$\;

\ForEach{(\texttt{comp\_epr}, \texttt{comp\_human}) in (\texttt{comps\_e}, \texttt{comps\_h})}{
    Initialize \texttt{l\_best} $\gets \infty$, \texttt{n\_best} $\gets 0$, \texttt{p\_best} $\gets -1$\;
    
    \ForEach{(\texttt{perf\_e}, \texttt{perf\_h}) in (\texttt{comp\_epr}, \texttt{comp\_human})}{
        \texttt{loss} $\gets$ MSE(\texttt{perf\_e['dur']}, \texttt{perf\_h['dur']})\;
        
        \If{\texttt{loss} $<$ \texttt{l\_best}}{
            \texttt{l\_best} $\gets$ \texttt{loss}\;
            \texttt{n\_best} $\gets$ len(\texttt{perf\_h['dur']})\;
        }
        
        \texttt{pearson} $\gets$ PearsonR(\texttt{perf\_e['dur']}, \texttt{perf\_h['dur']})\;
        
        \If{\texttt{pearson} $>$ \texttt{p\_best}}{
            \texttt{p\_best} $\gets$ \texttt{pearson}\;
        }
    }
    
    \texttt{l\_sum} $\gets$ \texttt{l\_sum} + (\texttt{l\_best} $\times$ \texttt{n\_best})\;
    \texttt{n\_notes} $\gets$ \texttt{n\_notes} + \texttt{n\_best}\;
    \texttt{p\_sum} $\gets$ \texttt{p\_sum} + \texttt{p\_best}\;
}

\texttt{l} $\gets$ \texttt{l\_sum} / \texttt{n\_notes}\;
\texttt{p} $\gets$ \texttt{p\_sum} / len(\texttt{comps\_e})\;

\Return{\texttt{l}, \texttt{p}}\;
\end{algorithm}

\section{Experiments}

\subsection{Experimental Setup}

For both the performance model and the discriminator, a learning rate of $1 \times 10^{-5}$ and batch size of 4 is used. The values used for the weight-balancing factors $\lambda_0$ to $\lambda_5$ are set to $[1, 3, 0.1, 0.1, 1, 1]$. Furthermore, the discriminator is trained every 5 epochs. For the feature extractor, the number of encoder blocks in the stack is 6. Each encoder block with the dimensionality of input and output $d_{\text{model}} = 256$, inner-layer the dimensionality $d_{ff} = 1024$, and the number of heads $h = 4$. The hyperparameters are chosen through coarse parameter search.

We employ early stopping to elicit three models with the best validation performance in terms of IOI, duration, and velocity features separately. By merging the predictions for the three features, we combine the models into Peransformer, an ensemble model.

Various Python libraries are used to process MIDI data \cite{raffel2014intuitive}, to build and train the model \cite{paszke2019pytorch}, and to calculate the Pearson correlation coefficient and perform hypothesis tests \cite{2020SciPy}.

\subsection{Ablation Study}\label{sec:ablation} 

\begin{table}[]
\centering
\resizebox{\columnwidth}{!}{%
\begin{tabular}{@{}lcccccc@{}}
\toprule
Model      & $l_{ioi}$ $\downarrow$      & $l_{dur}$ $\downarrow$      & $l_{vel}$ $\downarrow$      & $p_{ioi}$ $\uparrow$      & $p_{dur}$ $\uparrow$      & $p_{vel}$ $\uparrow$      \\ \midrule
Ensemble   & \textbf{0.0114} & 0.0896          & 220.47          & \textbf{0.9016} & \underline{0.7589}          & \underline{0.5122}          \\ \midrule
No $c$       & 0.0129          & 0.1770          & \underline{214.89}          & 0.8949          & 0.6230          & 0.4377          \\
No D pause & \underline{0.0116}          & 0.0994          & 342.72          & 0.8938          & 0.6336          & 0.3593          \\
No $\lambda$  & 0.0127          & \underline{0.0860}          & 253.06          & 0.8955          & 0.7568          & 0.4876          \\
No D score & 0.0125          & 0.1033          & 247.40          & \underline{0.8993}          & 0.7355          & 0.4841          \\
No MSE     & 0.0154          & 0.2012          & 352.42          & 0.8772          & 0.7074          & 0.3981          \\
No D       & 0.0119          & \textbf{0.0788} & \textbf{204.54} & 0.8972          & \textbf{0.7674} & \textbf{0.5361} \\ \bottomrule
\end{tabular}
}
\caption{Results of the ablation study. The best results are shown in \textbf{bold face} and the second-best results are marked with \underline{underline}.}
\label{tab:ablation}
\end{table}

To understand the role of various modules of the proposed model, and to facilitate later evaluation of the Generalized EPR Metrics (GEM), we perform an ablation study by applying Algorithm \ref{algo:metric} on the raw model output with z-score scaling removed, reproducing existing objective evaluation procedures for EPR models. The ablation models are: No $c$: set $c$ to 1 to remove loss balancing between compositions. No D pause: train the discriminator every epoch. No $\lambda$: set $\lambda$ to 1 to remove loss weighting. No D score: remove the score embedding from the discriminator to make it unconditional. No MSE: train with GAN loss only. No D: train with MSE loss only. The results are given in Table \ref{tab:ablation}.

Notably, the no D score model resembles the one proposed in \cite{renault2023expressive}, which also leverages unconditional GAN. However, a major difference remains between the two models: while the MSE loss in \cite{renault2023expressive} is calculated between the rendition and the score MIDI, our work calculates the loss between the rendition and the human performance.

Despite obtaining conspicuous results under objective evaluation, the no D model is arguably abusing the metrics by directly optimizing towards them, resulting in outputs that are "in the middle" of the distribution of human performances from the perspective of the metrics. While these outputs may have low losses, they are not necessarily inside of the distribution. Further inspections of the rendition samples also suggest that they are very similar to the input score MIDI file than to the target human performance, with less expressively varied local tempo, articulations, and dynamics compared to the ensemble model, corroborating the hypothesis.

Other ablation models exhibit various performance decays, evidenced by both automatic metrics results and inspections on rendition samples. More noticeable issues include unstable tempo (no $c$, no $\lambda$, no MSE), and sudden changes in dynamics (no D pause, no MSE). Overall, the ensemble model is the most well-rounded model among the ablation models.

\subsection{Generalized EPR Metrics (GEM)}

\begin{table}[]
\centering
\resizebox{\columnwidth}{!}{%
\begin{tabular}{@{}lcccccc@{}}
\toprule
 & $l_{ioi}$ $\uparrow$ & $l_{dur}$ $\uparrow$ & $l_{vel}$ $\uparrow$ & $p_{ioi}$ $\uparrow$ & $p_{dur}$ $\uparrow$  & $p_{vel}$ $\uparrow$ \\ \midrule
36 @ 99\% & 0.9296 & 0.8354 & 0.9971 & 0.7798 & 0.9695 & 0.9952 \\
All 180 & 0.9534 & 0.8702 & 0.9800 & 0.7401 & 0.8462 & 0.9918 \\ \bottomrule
\end{tabular}
}
\caption{Pearson correlation coefficients between the results of \ref{sec:ablation} and GEM.}
\label{tab:gemcorr}
\end{table}

To examine the reliability of GEM as a replacement for existing objective evaluation procedures, we reconduct the ablation study using GEM, and compare the results with the ones obtained in \ref{sec:ablation}. With the ensemble model being disassembled into the underlying IOI, duration, and velocity models, 180 result pairs are acquired from the 20 compositions of the test split using the 9 models being evaluated. The Pearson correlation coefficients between the results are shown in Table \ref{tab:gemcorr}. ``36 @ 99\%'' shows the results calculated from the 36 samples with more than 99\% of notes matched in the alignment, representing the performance of GEM at the highest alignment accuracy. ``All 180'' is calculated from all of the 180 samples, indicating the performance of GEM under typical conditions.

The results suggest that GEM make accurate evaluations. A very strong correlation with current metrics can be seen in the IOI loss, duration loss, velocity loss, duration Pearson, and velocity Pearson metrics, while a strong correlation is observed in the IOI Pearson metric.

GEM also show high reliability and remain robust against varying alignment accuracies, shown by the very similar correlation between GEM and current metrics on compositions where the alignment accuracy is high (36 @ 99\%) and typical (all 180).

\subsection{Quantitative Evaluation}\label{sec:quantitative}

\begin{table*}[]
\centering
\begin{tabular}{@{}llccccccc@{}}
\toprule
Model Type & Model & $l_{ioi}$ $\downarrow$ & $l_{dur}$ $\downarrow$ & $l_{vel}$ $\downarrow$ & $p_{ioi}$ $\uparrow$ & $p_{dur}$ $\uparrow$  & $p_{vel}$ $\uparrow$ & Match \% \\
\midrule
\multirow{3}{*}{Low-informed} & Ours & 0.0210 & 0.1528 & \textbf{233.91} & 0.8956 & \textbf{0.7802} & \textbf{0.5378} & 97.12 \\
& Renault \textit{et al.} \cite{renault2023expressive} & \textbf{0.0205} & 0.5291 & 2491.09 & \textbf{0.9006} & 0.5805 & 0.2668 & 96.61 \\
& Tang \textit{et al.} \cite{tang2023reconstructing} & 0.0550 & \textbf{0.1305} & 1022.75 & 0.5035 & 0.6124 & 0.0996 & 91.88 \\
\midrule
Highly-informed & Jeong \textit{et al.} \cite{jeong2019graph} & \textbf{0.0102} & \textbf{0.0710} & \textbf{136.51} & \textbf{0.9277} & \textbf{0.8057} & \textbf{0.6397} & 97.31 \\
\midrule
& Score & 0.0197 & 0.1768 & 798.82 & 0.9042 & 0.7897 & 0.3157 & 95.51 \\
& Human & \textbf{0.0071} & \textbf{0.0618} & 168.75 & \textbf{0.9419} & \textbf{0.8704} & \textbf{0.6761} & 96.34 \\
\bottomrule
\end{tabular}
\caption{Quantitative analysis with GEM. The best results up to the next horizontal rule are highlighted in \textbf{bold face}.}
\label{tab:quantitative}
\end{table*}

Taking advantage of the ability of GEM to evaluate any EPR model with output in the MIDI format, we conduct a quantitative evaluation on Peransformer and existing EPR models. The models being compared include two recent low-informed models \cite{renault2023expressive, tang2023reconstructing}, and one high-informed model \cite{jeong2019graph} which achieved state-of-the-art performance. On top of making direct comparisons among EPR models, we take one step further and apply GEM to the score MIDI files and human performances for additional context.

We adopt the procedure proposed in \cite{rajpurkar2016squad} to approximate human performance. The first human performance of each composition is taken as the surrogate of human expressiveness, and we evaluate it against the remaining human performances. As this procedure would require each composition to have at least 2 human performances, the quantitative evaluation was conducted on the subset of 13 compositions of the test dataset that meets this requirement. The results are shown in Table \ref{tab:quantitative}. ``Match \%'' shows the percentage of rendition notes successfully matched with the human performance.

In comparison, \cite{renault2023expressive} appears to struggle with the velocity predictions, possibly due to its tendency to produce high-velocity predictions. Among the renditions for the 20 compositions of the test split, 8 of which have the velocity clipped to the maximum value of 127 for every note in the composition. \cite{tang2023reconstructing} often greatly accelerates the composition. For example, the input score MIDI of the second movement of Piano Sonata No. 11 by Beethoven is 459 seconds long. However, the rendition was shrunk to only 76 seconds by \cite{tang2023reconstructing}, potentially harming its IOI metric results.

The results suggest that Peransformer performs significantly better in the velocity metrics while archiving better or similar results in the IOI and duration metrics against existing low-informed EPR models. Meanwhile, our model greatly narrows the gap to the highly-informed model from \cite{jeong2019graph}. The approximated human performance remains better than the EPR models.

\subsection{Subjective Evaluation}

\begin{figure}
\centering
\includegraphics[width=1\linewidth]{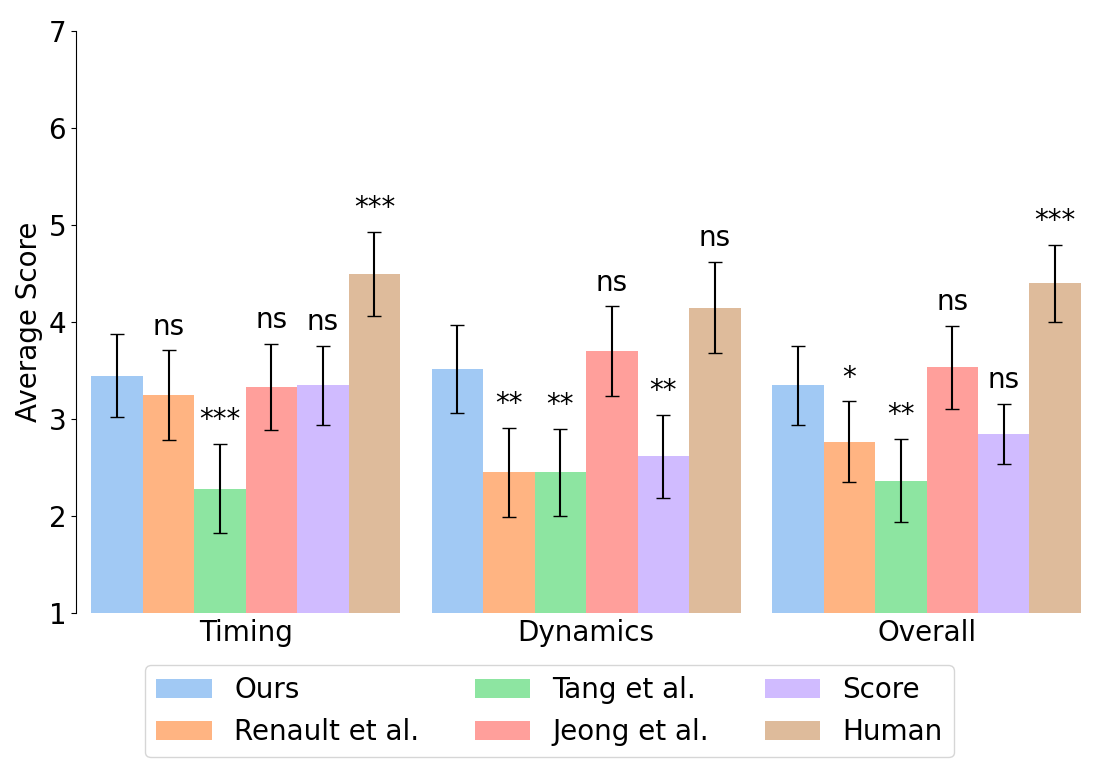}
\caption{Results of the subjective evaluation. The bar charts show the Mean Opinion Score (MOS) of the models. The error bars show the 95\% Confidence Interval (CI). The statistical significance of Welch's t-tests between Peransformer and the corresponding model is indicated above the error bars. ``ns'', ``*'', ``**'', and ``***'' denote $p \geq 0.05$, $p < 0.05$, $p < 0.01$, and $p < 0.001$ respectively.}
\label{fig:subjective}
\end{figure}

We conduct a survey to evaluate Peransformer and existing models qualitatively. 5 compositions with different compositional contents were selected from the test split, namely Beethoven's Piano Sonata No. 1 in F Minor, Op. 2 No. 1, Chopin's Etude in C major Op. 10 No. 7, Chopin's Ballade No. 2 in F major, Op. 38, Schubert's The Fantasie in C major, Op. 15, and Bach's Fugue in E-flat Major, BWV 876. The renditions are synthesized into audio clips using the same soundfont without pedaling. The audio clips are then trimmed to the first minute.

In this study, 60 participants aged over 21 were recruited. Among the participants, 51 (85\%) identified themselves as regular music listeners who listen to the music for more than 1 hour every week. 46 (76\%) claim to have at least 1 year of experience with some musical instruments.

The participants were asked to score each audio clip on a 7-point Likert scale regarding timing expressiveness, dynamics expressiveness, and overall expressiveness. Timing expressiveness evaluates the expressiveness of local tempo and articulation, relating to the IOI and duration predictions. Similarly, dynamics expressiveness is related to the velocity predictions and evaluates the expressiveness of dynamics. Finally, overall expressiveness is a holistic evaluation of the performance with all factors combined.

We discard responses that assign a higher overall expressiveness point to the score MIDI instead of the human performance as a quality control measure. The results are then aggregated across the compositions and shown in Figure \ref{fig:subjective}.

The qualitative results from the survey are fairly consistent with the quantitative results calculated using GEM. For timing, all models, unfortunately, show a similar level of expressiveness to the score MIDI files except for \cite{tang2023reconstructing}, which showed a lower performance. The human performances are still significantly more expressive in timing compared to the others.

In terms of dynamics, both Peransformer and the highly-informed \cite{jeong2019graph} showed high performances, achieving results similar to human performance, surpassing current low-informed models and the score MIDI files.

Overall, Peransformer significantly outperforms existing low-informed EPR models \cite{renault2023expressive, tang2023reconstructing} and has achieved results similar to the highly-informed model \cite{jeong2019graph}. The human performance remains the best across all metrics.

\subsection{Case Study: Comparison in Velocity}

\begin{figure}
\centering
\includegraphics[width=1\linewidth]{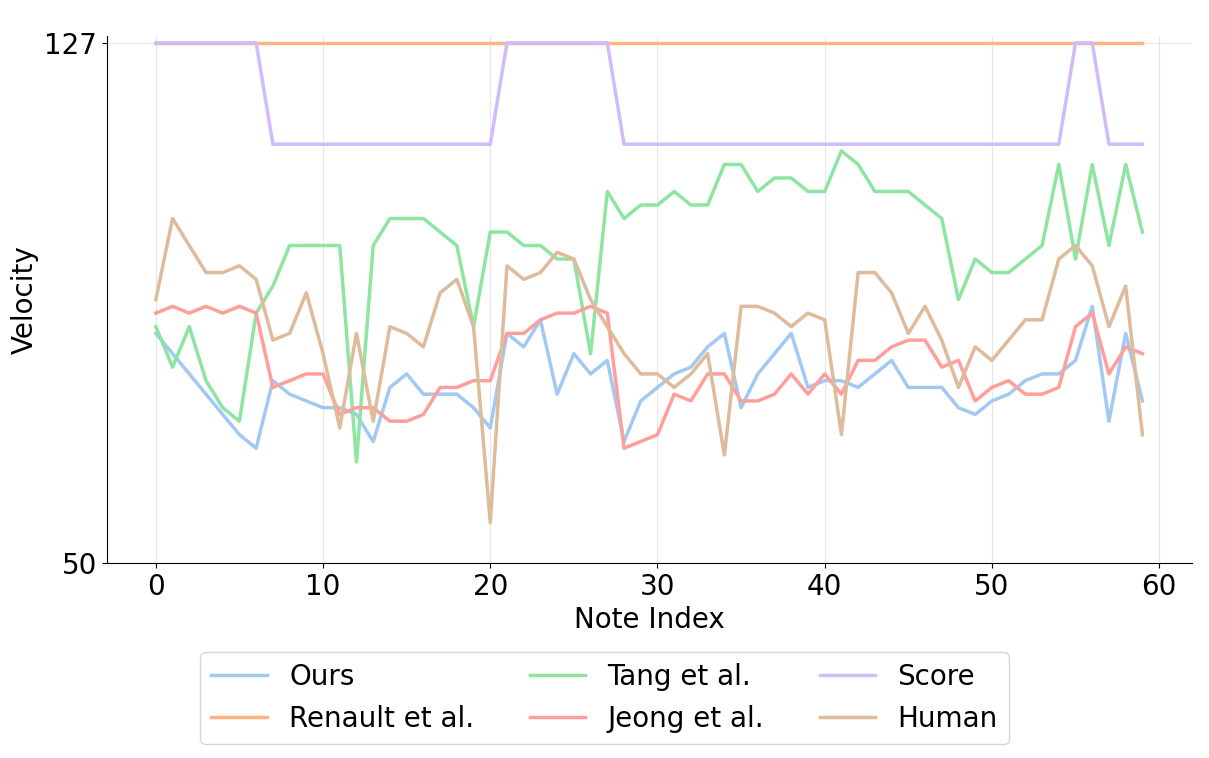}
\caption{Velocity changes of the first 60 notes of Schubert's The Fantasie in C major, Op. 15.}
\label{fig:velocity}
\end{figure}

To better understand the velocity predictions of the Peransformer model, which corresponds to the dynamics expressiveness in the subjective evaluation, we compare the velocity changes predicted by the EPR models along with the score MIDI and the human performance, using the first 60 notes of Schubert's The Fantasie in C major, Op. 15. The results are shown in Figure \ref{fig:velocity}.

The input score MIDI file arguably does not provide much meaningful information to assist the velocity prediction for the EPR models, as it only alternates between two values. The velocity predictions by \cite{renault2023expressive} are clipped to the maximum value of 127, showing a peculiar example of the observation discussed in \ref{sec:quantitative}. \cite{tang2023reconstructing} partially followed the human performance, while partially deviating from it.

Among the EPR models, the predictions of Peransformer and \cite{jeong2019graph} closely follow the human performance in general. Corroborating the observations from the quantitative and subjective analysis. Meanwhile, the human performance appears to have a more pronounced phrasing, indicated by the more obvious peaks and dips line chart, possibly adding extra expressiveness to the performance.

\section{Conclusion}

We proposed Peransformer, a low-informed EPR model trained with a score-aware discriminator and ASAP-MIDI - a score-to-performance paired, note-to-note aligned MIDI dataset. Quantitative and subjective evaluations show that Peransformer has achieved state-of-the-art results among current low-informed EPR models, while greatly narrowing the gap between low-informed and highly-informed EPR models. We also proposed the Generalized EPR Metrics (GEM), a standardized evaluation workflow that enables direct, accurate, and reliable evaluation of any EPR model with output in the MIDI format.

Future work can focus on further improving the performance of EPR models, especially on timing-related aspects, to match human expressiveness. The relation between the IOI, duration, and velocity metrics and the overall subjective expressiveness of performances can be further investigated to pave the road for the development of more comprehensive automatic metrics.

\printbibliography

\end{document}